\documentclass[aps,prl,twocolumn,superscriptaddress,showpacs,preprintnumbers,amsmath,amssymb]{revtex4-1}
\usepackage[dvips]{epsfig} 
\usepackage{dcolumn}  
\usepackage{color}  

\graphicspath{{ps}}

\begin{document}


\preprint{\vbox{ 
\hbox{Belle Preprint 2010-8   }
\hbox{KEK Preprint 2010-7}
}}
\vspace*{3\baselineskip}
\title{ \quad\\[1.0cm] Observation of $ B^+ \to \bar{D}^{*0} \tau^+ 
\nu_{\tau}$ 
and Evidence for $ B^+ \to \bar{D}^{0} \tau^+ 
\nu_{\tau}$ 
at Belle.}


\affiliation{Budker Institute of Nuclear Physics, Novosibirsk}
\affiliation{Faculty of Mathematics and Physics, Charles University, Prague}
\affiliation{Chiba University, Chiba}
\affiliation{University of Cincinnati, Cincinnati, Ohio 45221}
\affiliation{The Graduate University for Advanced Studies, Hayama}
\affiliation{Hanyang University, Seoul}
\affiliation{University of Hawaii, Honolulu, Hawaii 96822}
\affiliation{High Energy Accelerator Research Organization (KEK), Tsukuba}
\affiliation{Institute of High Energy Physics, Chinese Academy of Sciences, Beijing}
\affiliation{Institute of High Energy Physics, Vienna}
\affiliation{Institute of High Energy Physics, Protvino}
\affiliation{Institute of Mathematical Sciences, Chennai}
\affiliation{Institute for Theoretical and Experimental Physics, Moscow}
\affiliation{J. Stefan Institute, Ljubljana}
\affiliation{Kanagawa University, Yokohama}
\affiliation{Institut f\"ur Experimentelle Kernphysik, Karlsruher Institut f\"ur Technologie, Karlsruhe}
\affiliation{Korea Institute of Science and Technology Information, Daejeon}
\affiliation{Korea University, Seoul}
\affiliation{Kyungpook National University, Taegu}
\affiliation{\'Ecole Polytechnique F\'ed\'erale de Lausanne (EPFL), Lausanne}
\affiliation{Faculty of Mathematics and Physics, University of Ljubljana, Ljubljana}
\affiliation{University of Maribor, Maribor}
\affiliation{Max-Planck-Institut f\"ur Physik, M\"unchen}
\affiliation{University of Melbourne, School of Physics, Victoria 3010}
\affiliation{Nagoya University, Nagoya}
\affiliation{Nara Women's University, Nara}
\affiliation{National Central University, Chung-li}
\affiliation{National United University, Miao Li}
\affiliation{Department of Physics, National Taiwan University, Taipei}
\affiliation{H. Niewodniczanski Institute of Nuclear Physics, Krakow}
\affiliation{Nippon Dental University, Niigata}
\affiliation{Niigata University, Niigata}
\affiliation{Novosibirsk State University, Novosibirsk}
\affiliation{Osaka City University, Osaka}
\affiliation{Panjab University, Chandigarh}
\affiliation{Seoul National University, Seoul}
\affiliation{Shinshu University, Nagano}
\affiliation{Sungkyunkwan University, Suwon}
\affiliation{School of Physics, University of Sydney, NSW 2006}
\affiliation{Tata Institute of Fundamental Research, Mumbai}
\affiliation{Excellence Cluster Universe, Technische Universit\"at M\"unchen, Garching}
\affiliation{Toho University, Funabashi}
\affiliation{Tohoku Gakuin University, Tagajo}
\affiliation{Tohoku University, Sendai}
\affiliation{Department of Physics, University of Tokyo, Tokyo}
\affiliation{Tokyo Metropolitan University, Tokyo}
\affiliation{Tokyo University of Agriculture and Technology, Tokyo}
\affiliation{IPNAS, Virginia Polytechnic Institute and State University, Blacksburg, Virginia 24061}
\affiliation{Yonsei University, Seoul}
  \author{A.~Bozek}\affiliation{H. Niewodniczanski Institute of Nuclear Physics, Krakow} 
  \author{M.~Rozanska}\affiliation{H. Niewodniczanski Institute of Nuclear Physics, Krakow} 
  \author{I.~Adachi}\affiliation{High Energy Accelerator Research Organization (KEK), Tsukuba} 
  \author{H.~Aihara}\affiliation{Department of Physics, University of Tokyo, Tokyo} 
  \author{K.~Arinstein}\affiliation{Budker Institute of Nuclear Physics, Novosibirsk}\affiliation{Novosibirsk State University, Novosibirsk} 
  \author{V.~Aulchenko}\affiliation{Budker Institute of Nuclear Physics, Novosibirsk}\affiliation{Novosibirsk State University, Novosibirsk} 
  \author{T.~Aushev}\affiliation{\'Ecole Polytechnique F\'ed\'erale de Lausanne (EPFL), Lausanne}\affiliation{Institute for Theoretical and Experimental Physics, Moscow} 
  \author{T.~Aziz}\affiliation{Tata Institute of Fundamental Research, Mumbai} 
  \author{A.~M.~Bakich}\affiliation{School of Physics, University of Sydney, NSW 2006} 
  \author{V.~Bhardwaj}\affiliation{Panjab University, Chandigarh} 
% ...
\collaboration{The Belle Collaboration}

\begin{abstract}
We present measurements of
$B^+\to \bar{D}^{*0} \tau^+ \nu_{\tau}$ 
and
$B^+\to \bar{D}^{0} \tau^+ \nu_{\tau}$ 
decays in a 
data sample of $657 \times 10^6$ $B\bar{B}$ pairs
collected with the Belle detector at the KEKB 
asymmetric-energy $e^+e^-$ 
collider. 
We find $446^{+58}_{-56}$ events of the decay 
$B^+\to \bar{D}^{*0} \tau^+ \nu_{\tau}$ 
with a significance of 8.1 standard deviations, and
$146^{+42}_{-41}$ events of the decay 
$B^+\to \bar{D}^{0} \tau^+ \nu_{\tau}$
with a significance of 3.5 standard deviations. 
The latter signal provides the first
evidence for this decay mode.
The measured branching fractions are  $\mathcal{B}(B^+\to \bar{D}^{*0} \tau ^+ 
\nu_{\tau})=(2.12^{+0.28}_{-0.27} ({\rm stat}) \pm 0.29 ({\rm syst})) \% $
and  $\mathcal{B}(B^+\to \bar{D}^{0} \tau ^+
\nu_{\tau})=(0.77\pm 0.22 ({\rm stat}) \pm 0.12 ({\rm syst})) \% $.
\end{abstract}


\maketitle

\tighten

{\renewcommand{\thefootnote}{\fnsymbol{footnote}}}
\setcounter{footnote}{0}


Measurements of leptonic and semileptonic decays of $B$ mesons to the
$\tau$ lepton 
can provide important constraints on the Standard 
Model (SM) and its extensions. 
Due to the large mass of the lepton in the final state,  
these decays are sensitive 
probes of models with extended Higgs sectors \cite{Itoh}. 
Semileptonic modes with $b \to  c \tau^- {\bar\nu}_{\tau}$ 
\cite{CC} 
transitions provide more observables sensitive to new physics 
than purely leptonic
$B^+\to \tau^+ \nu_{\tau}$ decays.
Of particular interest is
$\tau$ polarization. 
The effects of new physics are expected
to be larger in $B\to \bar{D} \tau ^+ \nu_{\tau}$ than in
$B\to \bar{D}^* \tau ^+ \nu_{\tau}$.
We note that decays to the vector meson offer the interesting
possibility of studying correlations between the D* polarization
and other observables~\cite{Garisto}.

The  predicted branching fractions, based on the SM,
are around 1.4\% and 0.7\% 
for $B^0 \to {D}^{*-} \tau ^+ \nu_{\tau}$ 
and $B^0 \to {D}^- \tau ^+ \nu_{\tau}$, 
respectively (see {\it e.g.}, \cite{hwang}). 
Despite relatively large branching fractions, 
multiple neutrinos in the final states make the search for 
semi-tauonic $B$ decays very challenging. 
Inclusive and semi-inclusive branching fractions 
have been measured in LEP experiments \cite{lep}  
with  an average 
branching fraction of 
$\mathcal{B}(b \to  \tau \nu_{\tau} X)=(2.48\pm 
0.26)\%$ 
\cite{PDG}.  
The exclusive decay was first observed by Belle \cite{Matyja}
in the $B^0\to {D}^{*-} \tau ^+ \nu_{\tau}$ mode.
Other modes have also been measured by
BaBar \cite{BaBar-1} and
Belle \cite{Kozakai}. 
The results are still statistically limited.
In particular, the Belle prelimnary result~\cite{Kozakai} is the only evidence 
to date for  
$B^+\to \bar{D}^0 \tau ^+ \nu_{\tau}$.
Further improvements in precision could tightly constrain 
theoretical models.

Decays of $B$ mesons to multi-neutrino final states 
can be 
studied
at $B$-factories via the recoil 
of the accompanying 
$B$ meson ($B_{\rm tag}$).
Reconstruction of the 
$B_{\rm tag}$ allows one to calculate the missing four-momentum 
in the $B_{\rm sig}$ decay; this helps separate signal events
from copious backgrounds.
At the same time the presence of a $B_{\rm tag}$ strongly suppresses 
the combinatorial and continuum backgrounds.
The disadvantage is 
the low $B_{\rm tag}$ reconstruction efficiency.  
To increase statistics, we
 reconstruct the $B_{\rm tag}$ ``inclusively'' from all the 
remaining particles 
after the $B_{\rm sig}$ selection (see Ref.~\cite{Matyja}). 
A data sample consisting of $657 \times 10^6$ $B\bar{B}$ pairs is
used in this analysis.
It was collected with 
the Belle detector~\cite{Belle}
at the KEKB asymmetric-energy $e^+e^-$ (3.5 on 8 GeV) 
collider \cite{KEKB} operating at the $\Upsilon(4S)$ resonance 
($\sqrt{s}=10.58$ GeV). 


We use Monte Carlo (MC) simulations to estimate signal
efficiencies and background contributions.
Large samples of the signal 
$B^+\to \bar{D}^{(*)0} \tau ^+ \nu_{\tau}$ decays are generated with the 
EvtGen package \cite{evtgen} 
using the ISGW2 model \cite{isgw2}. Radiative effects are modeled
using the PHOTOS 
code \cite{photos}. 
We use large MC samples of continuum $q\bar{q}$ ($q=u,d,s,c$) 
and inclusive $B\bar{B}$ events to model the background.
The sizes of these samples are, respectively, six and nine times
that of the data.

Primary charged tracks 
are required to have
impact parameters  consistent with an origin at 
the interaction point 
(IP),
and to have momenta above 50 MeV/$c$ in the laboratory frame.
$K^0_S$ mesons are reconstructed using pairs of charged tracks 
 satisfying
$482 ~{\rm MeV}/c^2 < M_{\pi^+\pi^-} < 514 ~{\rm MeV}/c^2$
with 
a vertex displacement from the IP consistent with the
reconstructed momentum vector.
Muons, electrons, charged pions, kaons 
and protons are identified using information from particle 
identification subsystems \cite{PID}.  
The momenta of particles identified as electrons  are 
corrected for bremsstrahlung by adding photons within a 50 mrad cone 
along the lepton trajectory.

The  $\pi^0$ candidates are reconstructed from photon pairs 
having
118 MeV/$c^2<M_{\gamma\gamma}<150\ {\rm MeV}/c^2$. 
For candidates 
that share a common $\gamma$, we select the one with the 
smallest  $\chi^2$ value resulting from a $\pi ^0$ 
mass-constrained fit. 
To reduce the combinatorial background, we 
require that the photons from the $\pi^ 0$ have 
energies greater than 60 MeV - 120 MeV, 
depending on the photon's polar angle.
Photons that are not associated with a $\pi^0$ are accepted
if their energy  exceeds a polar-angle dependent threshold
ranging from 100 MeV to 200 MeV. 

The $\bar{D}^0$ candidates are reconstructed in the $K^+\pi^-$ and 
$K^+\pi^-\pi^0$ final states.
We accept
$\bar{D}^0$ 
candidates having an invariant mass in a 
$3\sigma$  window of the nominal $M_{D^0}$ mass.

The $\bar{D}^{*0}$ candidates are reconstructed from 
$\bar{D}^0 \pi^0$. 
We require that the mass difference
$\Delta M = M_{D^*}-M_{D^0}$ is in a 3$\sigma$  
window around its nominal value. 
We also accept $\bar{D}^0 \gamma$ pairs that do  
not fulfill the requirement on  $\Delta M$ if they are kinematically 
consistent with the hypothesis that
$\bar{D}^0$ and $\gamma$ 
come from the decay $\bar{D}^{*0}\to \bar{D}^0 \pi^0$ 
with one undetected photon ($\gamma_{\rm miss}$) from the $\pi^0$ decay 
(``partial reconstruction'' of $\bar{D}^{*0}$). 
For this purpose $\cos(\theta_{\gamma,\gamma_{\rm miss}})$, 
the cosine of the angle 
between two photons from the $\pi^0~$ is calculated in the $\bar{D}^{*0}$ rest 
frame  taking the nominal $\bar{D}^{*0}$ and $\pi^0$
masses.  
We require $|\cos(\theta_{\gamma,\gamma_{\rm miss}})|\:<\:1.1$ 
(taking into account experimental precision)
and that the energy of the detected photon 
exceeds 120 MeV.  
The partial reconstruction of $\bar{D}^{*0}$ increases
the reconstruction efficiency by a factor of about four, 
but due to higher background
it is only used in the subchannels  
with $\bar{D}^0 \to K^+ \pi^-$ decay.

To reconstruct the $\tau$ lepton candidates, we use
the $\tau^+ \to e^+\nu_e\bar{\nu}_{\tau}$, 
$\tau^+ \to \mu^+\nu_{\mu}\bar{\nu}_{\tau}$,  
and $\tau^+ \to \pi^+\bar{\nu}_{\tau}$ modes. 
In the latter case, we also take into account 
the contribution from the 
$\tau^+ \to \rho^+\bar{\nu}_{\tau}$ channel.
The $\tau^+\to \pi^+\bar{\nu}_{\tau}$ mode has a sensitivity 
similar to the  $\tau^+ \to e^+\nu_e\bar{\nu}_{\tau}$ or 
$\tau^+ \to \mu^+\nu_{\mu}\bar{\nu}_{\tau}$ mode, and
can be used to study $\tau$  polarization.  
For this channel, due to the higher combinatorial background, we analyze 
only 
the decay chains with the $\bar{D}^0 \to K^+ \pi^-$ mode.
In total, we consider 13 different decay chains, eight with $\bar{D}^{*0}$
and five with $\bar{D}^0$ in the final states.

The signal candidates are selected 
by combining a
$\bar{D}^{(*)0}$ meson  with an
appropriately charged electron, muon or pion.
In the sub-channels with the
$\tau^+ \to \pi^+\bar{\nu}_{\tau}$ decay, 
the large combinatorial background
is suppressed by  requiring the pion energy $E_{\pi} > 0.6$~GeV.
From multiple candidates we select a 
($\bar{D}^{(*)0}d^+_{\tau}$) pair
(throughout the paper $d_{\tau}$ stands for
the charged $\tau$ daughter: $e$, $\mu$ or $\pi$)
with the 
best ${\bar{D}^{(*)0}}$ candidate, 
based on the value of $\Delta M$ 
(for subchannels where $\Delta M$ is available) 
or $M_{D^0}$.
For the pairs sharing the same ${\bar{D}^{(*)0}}$ candidate,
we select the candidate with the largest 
 vertex probability fit on the tagging side.
 
Once a $B_{\rm sig}$ candidate is found, the remaining particles that 
are not assigned
to $B_{\rm sig}$
are 
used to reconstruct the $B_{\rm tag}$ decay.  The consistency of a 
$B_{\rm tag}$ 
candidate 
with a $B$-meson decay is checked using the beam-energy constrained mass and 
the energy difference variables:
$M_{\rm tag} = \sqrt{E^2_{\rm beam} - {\bf p}^2_{\rm tag}},~~
{\bf p}_{\rm tag} = \sum_i {\bf p}_i$,
and
$\Delta E_{\rm tag} = E_{\rm tag} - E_{\rm beam}, ~~ E_{\rm tag} 
= \sum_i E_i$,
where $E_{\rm beam}$ 
is the beam energy and ${\bf p}_i$ and $E_i$ 
denote the 3-momentum vector and energy of the $i$'th particle.
All quantities are evaluated in the $\Upsilon(4S)$ rest frame.
The summation is over all particles that are assigned 
to $B_{\rm tag}$.
We require that the candidate events have 
$M_{\rm tag}> 5.2 ~{\rm GeV}/c^2$
and $-0.3 ~{\rm GeV} <\Delta E_{\rm tag}<0.05 ~{\rm GeV}$.
With this requirement 
the  $M_{\rm tag}$ distribution of the signal 
peaks at the $B^+$ mass 
with about 80\% of the events being contained in the
signal-enhanced region
$M_{\rm tag}>$ 5.26 GeV/$c^2$. 

To suppress background and improve 
the quality of the $B_{\rm tag}$ selection, we impose the following 
requirements:
zero total event charge;     
no charged leptons in the event (except those coming from the signal side);
zero net proton/antiproton number; 
residual energy in the 
 electromagnetic calorimeter
({\it i.e.}, the sum of  energies 
that are not included in the $B_{\rm sig}$ nor $B_{\rm tag}$) 
should be less than 0.35 GeV (0.30 GeV or 0.25 GeV in sub-channels with higher 
backgrounds);
the number of neutral particles on the tagging side 
$N_{\pi^0}+N_{\gamma}<6$, $N_{\gamma} <3$,
and less than four tracks that do not satisfy
the requirements imposed on the impact parameters.
For decay modes with higher background, we impose further constraints 
on the total event strangeness and require
no $K^0_L$ in the event.
These criteria, which we refer to as 
``the $B_{\rm tag}$-selection'', reject events in 
which some particles were undetected and 
suppress events with a large number of spurious showers.  
In the samples of the ($\bar{D}^{(*)0}l^+$) pairs ($l=e, \mu$),
the dominant background comes from semileptonic $B$ decays,
$B^+\to \bar{D}^{(*)0} X l^+\nu_{l}$, whereas
in the case of the ($\bar{D}^{(*)0}\pi^+$) pairs,
the combinatorial background from hadronic $B$ decays dominates. 

Further background suppression exploits observables 
that characterize the signal decay: missing energy $E_{\rm miss} = 
E_{\rm beam}-E_{\bar{D}^{(*)0}}-E_{d_{\tau}^+}$; 
visible energy $E_{\rm vis}$, {\it i.e.}, 
the sum of the energies of all particles in the event; 
the square of missing mass 
$M_{\rm miss}^2 = E_{\rm miss}^2 - 
({\bf p}_{\rm sig} - 
{\bf p}_{\bar{D}^{(*)0}} -
{\bf p}_{d_{\tau}^+})^2$ 
and the effective mass of the 
($\tau^+ \nu_{\tau}$) pair, 
$q^2 = (E_{\rm beam} - 
E_{\bar{D}^{(*)0}})^2 - 
({\bf p}_{\rm sig} - 
{\bf p}_{\bar{D}^{(*)0}})^2$ 
where ${\bf p}_{\rm sig} = -{\bf p}_{\rm tag}$
(all kinematical variables are in the $\Upsilon(4S)$ 
rest frame).
The most useful 
variable for separating signal and background is 
obtained 
by combining $E_{\rm miss}$ and 
($\bar{D}^{(*)0} d_{\tau}^+$) pair momentum: 
$X_{\rm miss} = (E_{\rm miss} - |{\bf p}_{\bar{D}^{(*)0}} + 
{\bf p}_{d_{\tau}^+}|)/
\sqrt{E_{\rm beam}^2 -m_{B^+}^2}$ where $m_{B^+}$ is the $B^+$ mass.
The $X_{\rm miss}$ variable 
is closely related to
the missing mass in the $B_{\rm sig}$ decay but does not depend on 
the $B_{\rm tag}$ 
reconstruction \cite{Matyja}. 

The signal selection criteria are 
optimized individually in 
each decay chain, by maximizing
the expected significance $N_S/\sqrt{N_S+N_B}$, where $N_S$ 
and $N_B$
are the number of signal and background events in the signal-enhanced 
region,
assuming the SM 
prediction \cite{hwang} for the signal branching fractions.
The expected background $N_B$ is evaluated using generic MC samples. 
We require $E_{\rm vis}< 8.3 ~{\rm GeV}$ -- $8.5 ~{\rm GeV}$,  
$E_{\rm miss}> 1.5 ~{\rm GeV}$ -- $1.9 ~{\rm GeV}$
and  
$X_{\rm miss}>$ 2.0 -- 2.75 for leptonic $\tau$ decays,
or $X_{\rm miss}>$ 1.0 -- 1.5 for the modes with $\tau \to \pi \nu_{\tau}$.
In the latter case, where the $\tau$ decays 
to a final state with a single neutrino, 
we further require
$\cos(\theta_{\nu_1\nu_2})$ to be in the range 
$[-1,1]$, 
where $\theta_{\nu_1\nu_2}$ 
denotes the angle between the two neutrinos in the 
($\tau^+\nu_{\tau}$) rest frame
and is calculated from the $M_{\rm miss}^2$ and 
$q^2$ variables. 
In the sample with ($\bar{D}^0d^+_{\tau}$) pairs,
to suppress the cross-feeds from the  
$B\to \bar{D}^*\tau^+ \nu_{\tau}$ modes,
we impose a loose requirement on $q^2< 9.5 ~{\rm GeV}^2/c^4$.
 
The above 
requirements result in flat $M_{\rm tag}$ distributions for 
most background components, 
while the distribution of the signal modes remains unchanged. 
The main sources of the peaking 
background are
the semileptonic decays 
$B^+ \to \bar{D}^{*0}l^+\nu_l$
and $B \to \bar{D}^{(*)}\pi l^+\nu_l$
(including $\bar{D}^{**}l^+\nu_l$).

In order to estimate the peaking background reliably,
in particular from poorly known 
semileptonic modes of the type 
$B\to \bar{D}^{**}l\nu_l$, 
we divide the MC sample 
into the following categories:
$B\to\bar{D}^*l^+\nu_l$, 
$B\to\bar{D}l^+\nu_l$, 
$B\to\bar{D}^{**}l^+\nu_l$, 
other $B$ decays, $c\bar{c}$ and 
($u\bar{u}+d\bar{d}+s\bar{s}$) continuum.
The normalizations of these components are determined 
from simultaneous fits to experimental distributions of
$M_{\rm tag}$, $\Delta E_{\rm tag}$, $E_{d_{\tau}}$, $X_{\rm miss}$, 
$E_{\rm vis}$, $q^2$, 
and $R_2$, the ratio of the second and zeroth Fox-Wolfram
moments~\cite{FW}.
These fits are performed separately for the subsamples 
defined by the
($\bar{D}^{(*)0}d^+_{\tau}$) 
pairs, excluding the region  
$M_{\rm tag}>5.26 ~{\rm GeV/c}^2$ and $X_{\rm miss}>2.0$, 
where we expect enhanced signal contributions. 

The signal and combinatorial background yields are extracted from an extended
unbinned maximum likelihood fit to the
$M_{\rm tag}$ and $P_{D^0}$ (momentum of $D^0$ from $B_{\rm sig}$
measured in the $\Upsilon(4S)$ frame) variables.  
The $M_{\rm tag}$ variable allows us to
separate the combinatorial background from the signal, while $P_{D^0}$
helps to distinguish between the two signal modes.  Correlations
between these variables are found to be small.

 Parameterizations of two-dimensional probability density
  functions (PDFs) are determined from the MC samples.  They are
  expressed as the product of one-dimensional PDF's for each variable.
The PDF's for $M_{\rm tag}$ of the signal and peaking 
background components are  described using 
an empirical parameterization introduced by the Crystal Ball 
collaboration \cite{CB}, while 
combinatorial backgrounds are parameterized by the ARGUS function
\cite{ARGUS}.
It has been empirically found that the PDF's for $P_{D^0}$ are 
well modeled as a sum of two  Gaussian distributions.

 The fits are
performed in the range $M_{\rm tag}>5.2$ GeV/$c^2$, 
simultaneously to all data subsets.
In each of the subchannels, we describe the data as the sum of
four components: signal, cross-feed between  
$\bar{D}^{*0}\tau^+\nu_{\tau}$ and $\bar{D}^{0}\tau^+\nu_{\tau}$, 
combinatorial and peaking 
backgrounds.
The common signal branching fractions
$\mathcal{B}(B^+\to \bar{D}^{*0}\tau^+\nu_{\tau})$
and $\mathcal{B}(B^+\to \bar{D}^{0}\tau^+\nu_{\tau})$,
and the numbers of combinatorial background in each subchannel 
are free parameters of the fit,
while the normalizations of peaking background contributions are 
fixed 
to the values 
obtained 
from the rescaled MC samples 
(as described above). 
The signal yields and branching fractions for 
$B^+\to \bar{D}^{(*)0}\tau^+\nu_{\tau}$ decays are
related using the following 
formula, which assumes
equal fractions of charged and 
neutral $B$ meson pairs produced in $\Upsilon(4S)$ decays:
$\mathcal{B}(B^+\to\bar{D}^{(*)0}\tau^+\nu_{\tau}) = 
N_s^{D^{(*)}}/(N_{B\bar{B}}\times \sum_{k}\epsilon_{k}\mathcal{B}_{k})$,
where $N_{B\bar{B}}$ is the number of 
$B\bar{B}$ pairs and the index $k$ runs over the 13 decay chains;
$\epsilon_{k}$ 
denotes the reconstruction 
efficiency of the  specific subchannel and  $\mathcal{B}_{k}$ is the 
product 
of intermediate branching 
fractions.
All the intermediate branching fractions are taken from the 
PDG compilation \cite{PDG}. 
The efficiencies of the signal reconstruction,  
as well as
the expected combinatorial and peaking 
backgrounds 
are given 
in Table~\ref{tab-yields}. 

\begin{figure}[htb]
\includegraphics[width=0.22\textwidth]{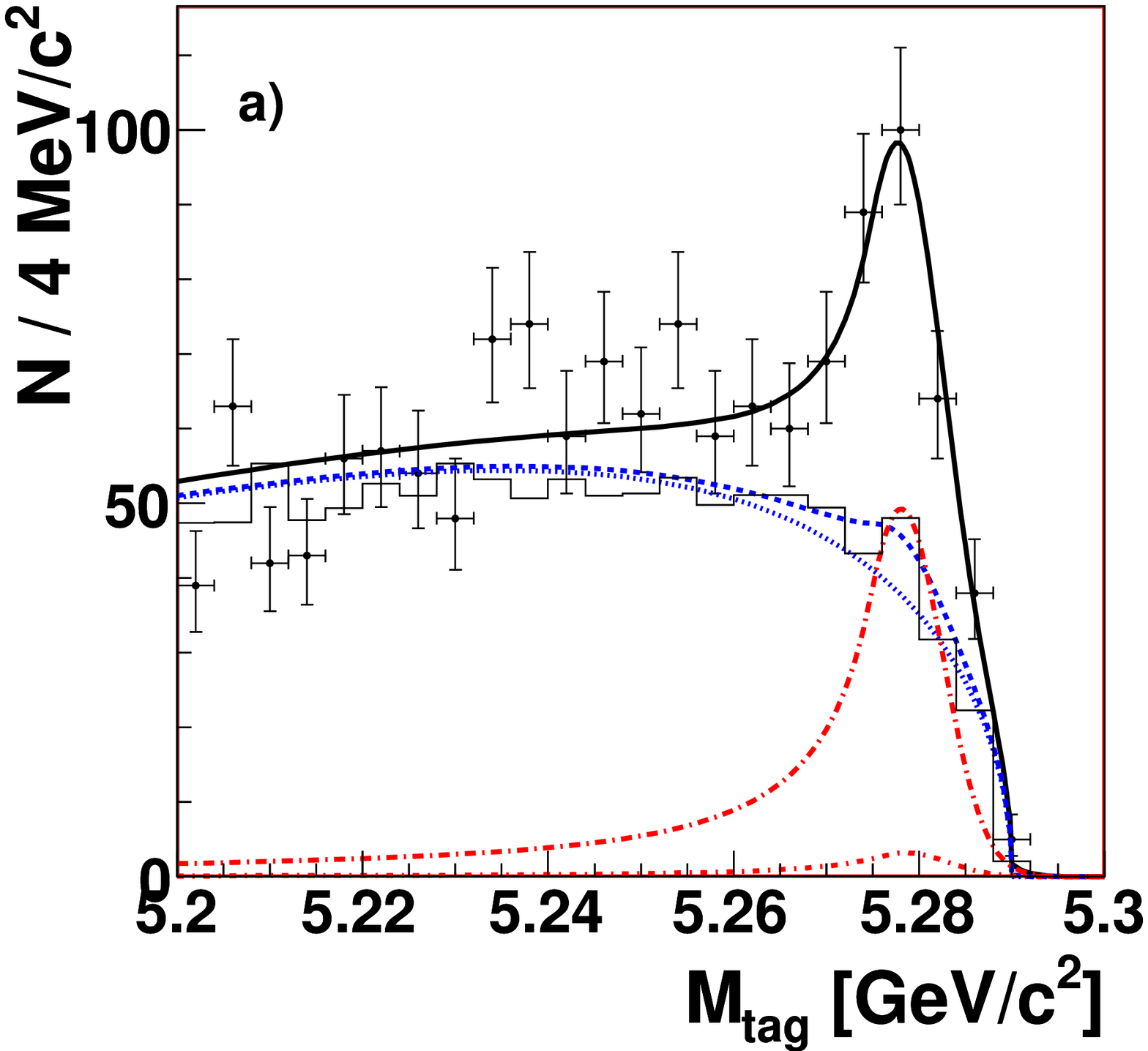}
\includegraphics[width=0.22\textwidth]{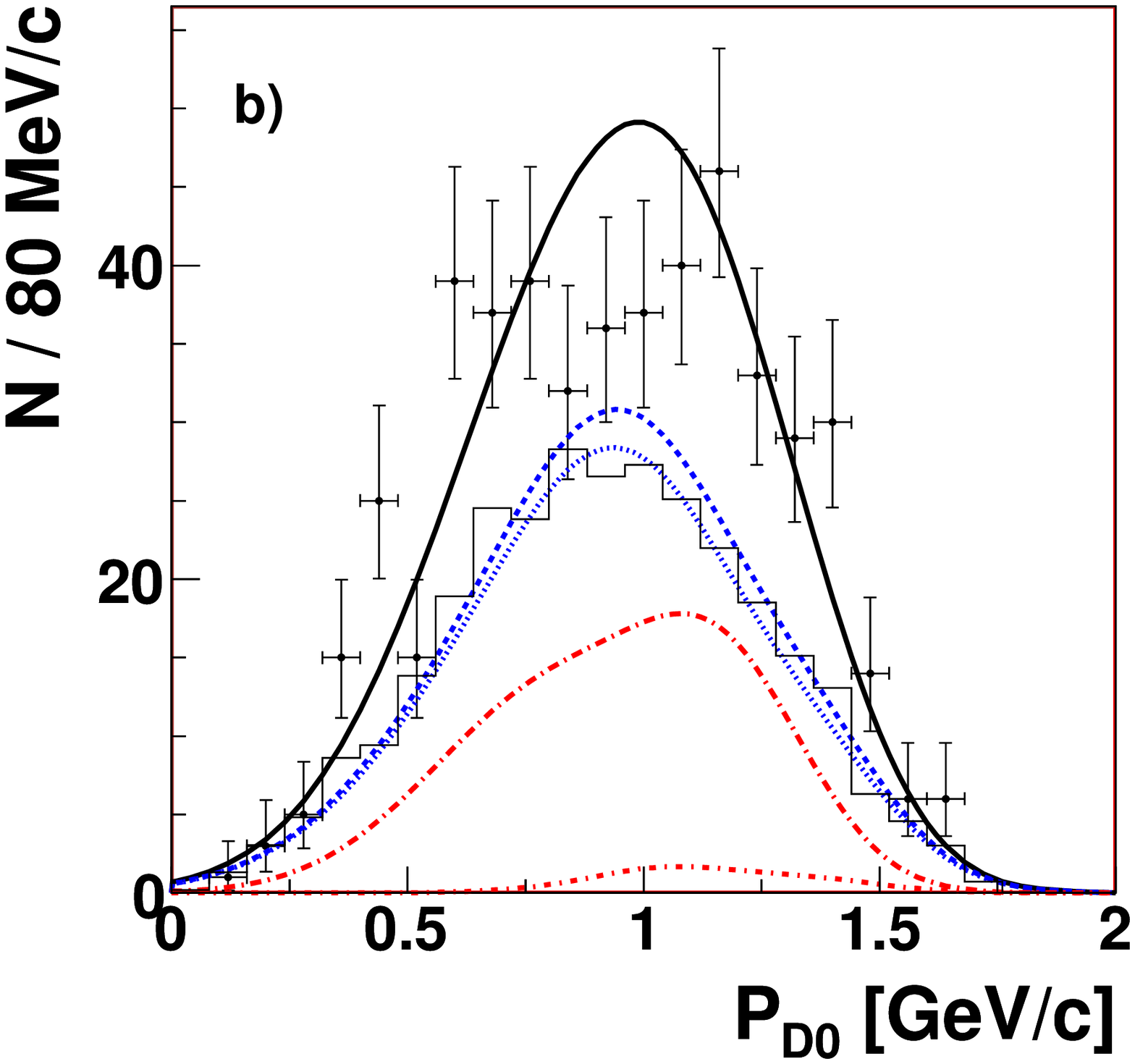}
\includegraphics[width=0.22\textwidth]{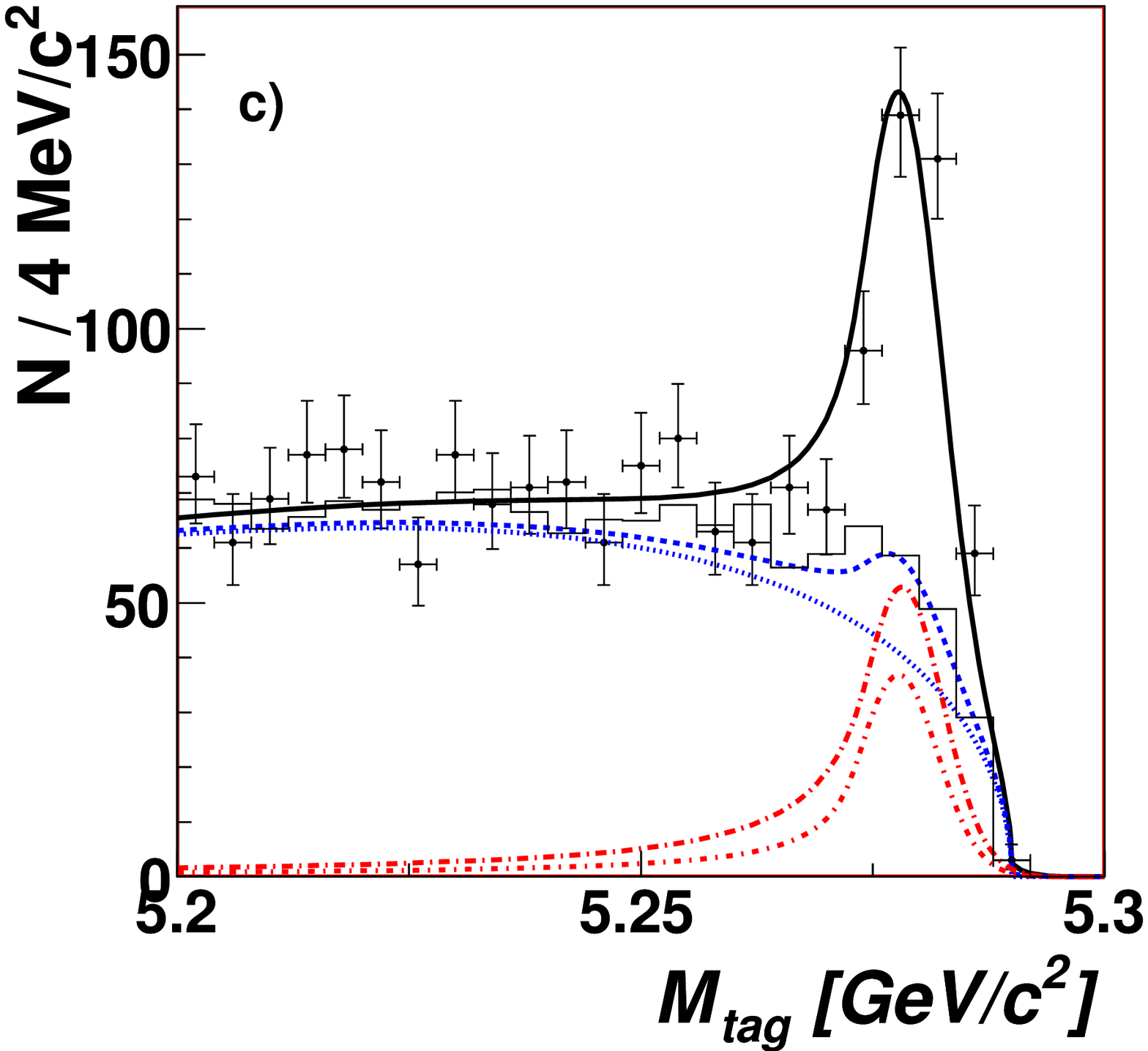}
\includegraphics[width=0.22\textwidth]{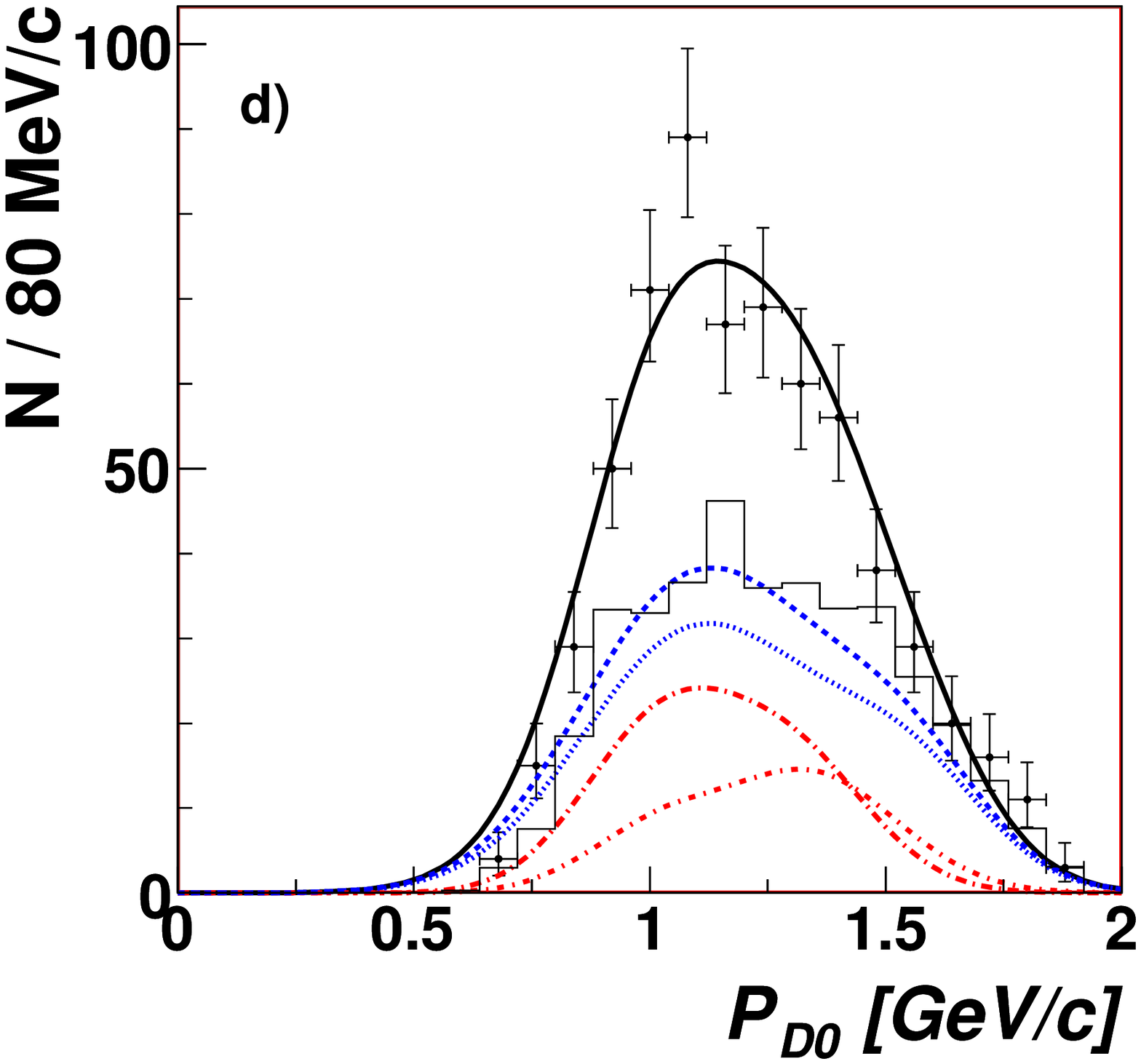}
\caption{The fit projections to
$M_{\rm tag}$,
and $P_{D^0}$ for $M_{\rm tag}>5.26 ~{\rm GeV}/c^2$  
(a,b) for $\bar{D}^{*0}\tau^+\nu_{\tau}$, (c,d)
for $\bar{D}^{0}\tau^+\nu_{\tau}$.
The black curves show the result of the fits. 
The solid dashed curves represent the background and 
the dashed dotted  ones show the combinatorial component. 
The dot-long-dashed and dot-short-dashed curvess represent, respectively, the signal 
contributions from
$B^+\to\bar{D}^{*0}\tau^+\nu_{\tau}$ 
and $B^+\to\bar{D}^{0}\tau^+\nu_{\tau}$.
The histograms represent the MC-predicted background.
}
\label{pic-fit}
\end{figure}

The signal extraction procedure has been tested 
by fitting ensembles of simulated experiments
containing all signal and background components. 
These pseudo-experiments are generated
using the shapes of the fitted PDF's for 
the signal and background components 
and with the number of events are Poisson-distributed
around the expected yields.
The pull distributions of the extracted signal branching
fractions are consistent with standard normal distributions.
The small biases in the mean values 
are included in the final systematic 
uncertainties.
\begin{table*}[hbt]
\caption{The yields of signal ($N_s$) 
and combinatorial background ($N_b$) events determined
from fits to data,  
number of expected 
combinatorial ($N_b^{\rm MC}$) 
and peaking  ($N_p$)
background events,
signal selection efficiencies ($\epsilon=\sum_{k}\epsilon_{k}\mathcal{B}_{k}$),
extracted branching fractions 
($\mathcal{B}$)
and statistical significances ($\Sigma$).
$N_p$ and $N_b^{\rm MC}$ are evaluated from fits of the generic 
MC samples to experimental distributions.  
The numbers in parentheses refer to the signal 
$\bar{D}^{*0}(\bar{D}^0)$ modes
reconstructed as 
$\bar{D}^0(\bar{D}^{*0})$.
The efficiencies include intermediate branching fractions. 
The listed errors are statistical only.
The results are summed
over the considered $\bar{D}^{(*)0}$ and $\tau$ decay modes.
}
\begin{tabular}{ l c c c c c c c c c}
\hline
Mode & $N_{s}$
&$N_b$
&$N_b^{\rm MC}$
& $N_p$
&$ \epsilon (10^{-6})$ 
& $\mathcal{B}(\%)$
& $\Sigma (\sigma)$\\
\hline 
\hline 
$\bar{D}^{*0}\tau^+\nu_{\tau}$ 
& $446^{+58}_{-56} (226)$ 
& $1075^{+37}_{-35}$
& $1029\pm 20$
& $31.0\pm 17.7$ 
& $32.6\pm 0.2 (16.3)$
& $2.12^{+0.28}_{-0.27}$
& $8.8$\\
$\bar{D}^{0}\tau^+\nu_{\tau}$ 
& $146^{+42}_{-41} (15)$ 
& $1245^{+40}_{-39}$
& $1310\pm 19$
& $78.2\pm 12.6 $
& $30.0\pm 0.4 (3.2)$
& $0.77\pm 0.22$
& $3.6$ \\
\hline
\hline
\end{tabular}
\label{tab-yields}
\end{table*}
The procedure 
established above 
is applied to the 
data. 
The $M_{\rm tag}$ and $P_{D^0}$ distributions
for the $\bar{D}^{*0}\tau^+\nu_{\tau}$ and   
$\bar{D}^{0}\tau^+\nu_{\tau}$ samples in data   
are shown in Fig.~\ref{pic-fit}.
The overlaid histograms represent the expected background, scaled to the 
data luminosity. A clear excess of events over  
background is visible in the signal-enhanced region. 

The branching fractions extracted from the fit are 
$\mathcal{B}(B^+\to \bar{D}^{*0} \tau ^+ 
\nu_{\tau})=(2.12^{+0.28}_{-0.27} ({\rm stat})) \% $
and  $\mathcal{B}(B^+\to \bar{D}^{0} \tau ^+
\nu_{\tau})=(0.77 \pm 0.22 ({\rm stat})) \% $.
The signal yields are 
$446^{+58}_{-56}$ 
$B^+\to \bar{D}^{*0} \tau^+ \nu_{\tau}$ events
and $146^{+42}_{-41}$ 
$B^+\to \bar{D}^{0} \tau^+ \nu_{\tau}$ events.
The statistical significances, defined as $\Sigma = 
\sqrt{-2{\ln}(\mathcal{L}_{\rm 0}/\mathcal{L}_{\rm max})}$, corespond to
 8.8 and 3.6 standard deviations ($\sigma$), respectively.
Here
$\mathcal{L}_{\rm max}$ 
denotes the maximum likelihood value 
and $\mathcal{L}_{\rm 0}$ is the likelihood
for the zero signal hypothesis.
The fitted yields of combinatorial background in the individual submodes
are consistent within statistical uncertainties with the MC-based
expectations. 
The fit results are summarized in Table~\ref{tab-yields}. 
The fit projections in $M_{\rm tag}$ and $P_{D^0}$ 
are shown in Fig.~\ref{pic-fit}.

As a cross-check, we extract the signal yields
from an extended unbinned maximum likelihood fit to one-dimensional
distributions in $M_{\rm tag}$ and obtain consistent results
with the two-dimensional fit.
We also examine the distributions of variables 
used in the signal selection, applying all requirements
except 
those that are related to the considered variable. In all cases the 
distributions are well-reproduced by the sum of signal and 
background components with normalizations fixed from the fit to 
the $(M_{\rm tag},P_{D^0})$ distribution.
   
The systematic uncertainties in
the branching fractions are summarized in
Table~\ref{tab-sys}. 
They include uncertainties 
in the total number of $B\bar{B}$ pairs, 
the effective efficiencies
$\sum_{k}\epsilon_{k}\mathcal{B}_{k}$,
and the signal-yield extractions.
The systematic uncertainties associated with the effective efficiencies  
include errors in determination of the efficiencies for 
$B_{\rm tag}$ 
reconstruction and
($\bar{D}^{(*)0}d^+_{\tau}$) pair 
selection, coming from efficiencies of tracking,
neutral particle reconstruction,
particle identification,
and from imperfect modeling of real processes.
The uncertainty in the $B_{\rm tag}$ 
and part of the $B_{\rm sig}$ reconstruction efficiency is evaluated
from data control samples with 
$B^+\to\bar{D}^{*0}\pi^+$ and 
$B^+\to\bar{D}^{0}\pi^+$ decays on the signal-side.
The absolute normalizations of the data and MC control samples agree 
to within 13\%.  The difference, as well as uncertainties in 
the relative amounts of $D^{*0}-D^{0}$ cross-feeds are included in the 
systematic uncertainty of $B_{\rm tag}$ and $B_{\rm sig}$ reconstruction. 
The remaining uncertainties
in the lepton identification and signal selection are estimated
separately. The latter are determined
 by comparing MC and data distributions in the 
variables used for signal 
selection. 
The uncertainties due to the partial branching fractions $\mathcal{B}_k$ 
are taken from the errors quoted by the PDG \cite{PDG}.  

The systematic uncertainties in the signal yield originate from the background
evaluation and from the PDF parameterizations of the signal and
background components. 
The resulting error is evaluated from changes
in the signal yields obtained from fits where 
the PDF parameters and the relative contributions 
of the background
components are varied by  $\pm 1\sigma$.

All of the 
above
sources of systematic uncertainties are combined together taking into 
account correlations between different decay chains.  The combined 
systematic uncertainty is 13.9\% for the 
$B^+ \to \bar{D}^{*0}\tau^+\nu_{\tau}$ mode
and 15.2\%
for $B^+\to\bar{D}^{0}\tau^+\nu_{\tau}$.
\begin{table}[hbt]
\caption{Summary of the systematic uncertainties.} 
\begin{tabular}{ l c c }
\hline
Source
&$\bar{D}^{*0}\tau^+\nu_{\tau}$
&$\bar{D}^{0}\tau^+\nu_{\tau}$
\\
\hline \hline
$N_{B\bar{B}}$ & $\pm 1.4$\% & $\pm 1.4$\% \\
 Reconstruction of $B_{\rm tag}$ and $B_{\rm sig}$ & $\pm 12.9$\% & $\pm 12.8$\% \\
 Lepton-id and signal selection & $^{+1.5}_{-1.6}$\% & 
$^{+4.4}_{-4.5}$\% \\
Shape of the signal PDF's & $\pm 2.5$\% & $\pm 6.0$\% \\
Comb. and peaking backgrounds & $\pm 3.3$\% & $\pm 2.7$\% \\
Fitting procedure & $\pm 0.8$\% & $\pm 1.5$\% \\
\hline
 Total & $\pm 13.9$\% & $\pm 15.2$\% \\
\hline \hline\end{tabular}
\label{tab-sys}
\end{table}

We include the effect of systematic uncertainties in the signal 
yields on the significances of the observed signals by convolving the 
likelihood function from the fit  with a Gaussian systematic error 
distribution. The significances of the observed signals after including 
systematic uncertainties are 
8.1$\sigma$ and 3.5$\sigma$ for the $B^+\to \bar{D}^{*0} \tau^+ \nu_{\tau}$ 
and $B^+\to \bar{D}^{0} \tau^+ \nu_{\tau}$ modes, respectively.

In conclusion, in a sample of  
657$\times 10^6~B\bar{B}$ pairs 
we 
measure branching fractions 
$\mathcal{B}(B^+\to \bar{D}^{*0}\tau^+\nu_{\tau}) = (2.12^{+0.28}_{-0.27} 
({\rm stat}) \pm 0.29({\rm syst}))$\%, and
$\mathcal{B}(B^+\to \bar{D}^{0}\tau^+\nu_{\tau}) = (0.77 \pm 0.22 
({\rm stat}) \pm 0.12({\rm syst}))$\%, 
which are consistent  within experimental uncertainties with SM 
expectations \cite{hwang}.
The result on $B^+\to \bar{D}^0\tau ^+\nu_{\tau}$ 
is the first evidence for this decay mode. 

We thank the KEKB group for excellent operation of the
accelerator, the KEK cryogenics group for efficient solenoid
operations, and the KEK computer group and
the NII for valuable computing and SINET3 network support.  
We acknowledge support from MEXT, JSPS and Nagoya's TLPRC (Japan);
ARC and DIISR (Australia); NSFC (China); MSMT (Czechia);
DST (India); MEST, NRF, NSDC of KISTI, and WCU (Korea); MNiSW (Poland); 
MES and RFAAE (Russia); ARRS (Slovenia); SNSF (Switzerland); 
NSC and MOE (Taiwan); and DOE (USA).


%

\begin{thebibliography}{99}

\bibitem{Itoh} A.~S.~Cornell {\it et al.}, 
arXiv:0906.1652 [hep-ph] and references quoted therein.

\bibitem{CC}
Throughout this paper, 
the inclusion of the charge-conjugate decay mode is implied
unless otherwise stated.

\bibitem{Garisto}
R.~Garisto, 
Phys. Rev. D {\bf 51}, 1107 (1995);
M.~Tanaka, Z. Phys. C {\bf 67}, 321 (1995). 
\bibitem{hwang}
C.-H.~Chen and C.-Q.~Geng, JHEP {0610}, 053 (2006).  

\bibitem{lep}
G.~Abbiendi {\it et al.} (OPAL Collaboration), Phys. Lett. B {\bf 520}, 
1 (2001);
R.~Barate {\it et al.} (ALEPH Collaboration), Eur. Phys. J. C  {\bf 19}, 
213
(1996);
P. Abreu {\it et al.} (DELPHI Collaboration), Phys. Lett. B {\bf 496}, 
43 
(2000);
M.~Acciarri {\it et al.} (L3 Collaboration), Z. Phys. C {\bf 71}, 379 
(1996).
\bibitem{PDG}
C. Amsler {\it et al.} (Particle Data Group), Phys. Lett. B {\bf 667}, 1 
(2008).
\bibitem{Matyja}
A.~Matyja {\it et al.} (Belle Collaboration), Phys. Rev. Lett. {\bf 
99}, 191807 (2007).

\bibitem{BaBar-1}
B. Aubert {\it et al.} (BaBar Collaboration), Phys. Rev. Lett. {\bf 
100}, 021801 (2008).

\bibitem{Kozakai}
I. Adachi {\it et al.} (Belle Collaboration), BELLE-CONF-0901, 
arXiv:0910.4301 [hep-ex]. 

\bibitem{Belle}
A.~Abashian {\it et al.} (Belle Collaboration),
Nucl. Instr. and Meth. A {\bf 479}, 117 (2002).

\bibitem{KEKB}
S.~Kurokawa and E.~Kikutani, Nucl. Instr. and. Meth. A {\bf 499}, 1 
(2003),
and other papers included in this volume.

\bibitem{evtgen} 
D.~J.~Lange, 
Nucl. Instr. and Meth. A {\bf 462}, 152 (2001).
\bibitem{isgw2}D.~Scora and N.~Isgur, Phys. Rev. D {\bf 52}, 2783 
(1995).
\bibitem{photos}E.~Barberio and Z.~W\c{a}s, Comput. Phys. Commun. {\bf 79}, 
291 
(1994).
\bibitem{PID}
E.~Nakano {\it et al.}, Nucl. Instr. Meth. A {\bf 494}, 402 (2002).
\bibitem{FW}
 G.~C.~Fox and S.~Wolfram, Phys. Rev. Lett. {\bf 41}, 1581 (1978).

%
%
%
%

\bibitem{CB}
T.~Skwarnicki, Ph.D.~Thesis, Institute of Nuclear Physics, Krakow 1986; 
DESY Internal Report, DESY F31-86-02 (1986).
\bibitem{ARGUS}H.~Albrecht et al. (ARGUS Collaboration), Phys. Lett. B 
{\bf 241}, 278 (1990).
\end{thebibliography}
\end{document}